\documentclass[twocolumn,pra,aps,showpacs,superscriptaddress,floatfix,showkeys,nofootinbib,10pt]{revtex4-2}

\usepackage[T1]{fontenc}
\usepackage{amsmath,amsfonts,amssymb,amsthm}
\usepackage{graphicx}

\newcommand{\ket}[1]{ | \, #1 \rangle}

\newcommand{\be}{\begin{equation}} \newcommand{\ee}{\end{equation}}
\newcommand{\ba}{\begin{aligned}} \newcommand{\ea}{\end{aligned}}

\DeclareRobustCommand\openone{\leavevmode\hbox{\small1\normalsize\kern-.33em1}}%

\begin{document}

\title{Optimization of experimental quantum randomness expansion}
\author{Amelie Piveteau}
\affiliation{Department of Physics, Stockholm University, S-10691 Stockholm, Sweden}
\author{Alban Seguinard}
\affiliation{Department of Physics, Stockholm University, S-10691 Stockholm, Sweden}
\author{Piotr Mironowicz} \email{piotr.mironowicz@gmail.com}
\affiliation{Department of Physics, Stockholm University, S-10691 Stockholm, Sweden}
\affiliation{Department of Algorithms and System Modeling, Faculty of Electronics, Telecommunications and Informatics, Gda\'{n}sk University of Technology, Poland}
%\affiliation{International Centre for Theory of Quantum Technologies, University of Gda\'{n}sk, Wita Stwosza 63, 80-308 Gda\'{n}sk, Poland}
%\affiliation{Department of Physics, Stockholm University, S-10691 Stockholm, Sweden} 
%\affiliation{National Quantum Information Centre in Gda\'nsk, 81-824 Sopot, Poland}
%\affiliation{Faculty of Applied Physics and Mathematics, Gda\'{n}sk University of Technology, Gabriela Narutowicza 11/12, 80-233 Gda\'{n}sk, Poland}
%\affiliation{Institute of Theoretical Physics and Astrophysics, University of Gda\'{n}sk, Wita Stwosza 63, 80-308 Gda\'{n}sk, Poland}
\author{Mohamed Bourennane}
\affiliation{Department of Physics, Stockholm University, S-10691 Stockholm, Sweden}

\date{\today}

\begin{abstract}
	Quantum technologies provide many applications for information processing tasks that are impossible to realize within classical physics. These capabilities include such fundamental resources as generating secure, i.e. private and unpredictable random values. Yet, the problem of quantifying the amount of generated randomness is still not fully solved.
	This work presents a comprehensive analysis of the design and performance optimization of a Quantum Random Number Generator (QRNG) based on Bell inequality violations. We investigate key protocol parameters, including the smoothing parameter ($\epsilon_{\text{s}}$), test round probability ($\gamma$), and switching delays, and their effects on the generation rate and quality of randomness. We identify optimal ranges for $\gamma$ and $p_\Omega$ (the protocol's non-aborting probability) to balance the trade-off between randomness consumption and net randomness generation. Additionally, we explore the impact of switching delays on the system's performance, providing strategies to mitigate these effects. Our results indicate substantial developments in QRNG implementations and offer higher randomness expansion rates. The work provides practical guidelines for the efficient and secure design of QRNG systems and other cryptographic protocols.
\end{abstract}

%\keywords{quantum}

\maketitle

	Quantum randomness certification is a crucial aspect of quantum information theory, particularly in applications such as cryptography, secure communications, and random number generation~\cite{pironio2010random,acin2016certified,liu2018device,liu2021device}. In classical systems, randomness can be generated using physical processes or algorithms; however, ensuring that this randomness is truly unpredictable and not biased or manipulated is challenging, especially when the devices are not trusted~\cite{mayers2003self}. Quantum mechanics offers a unique approach to randomness through phenomena like superposition and entanglement~\cite{horodecki2009quantum}.

Device-independent quantum protocols are designed to operate without trusting the internal workings of the devices used for generating random numbers~\cite{renner2008security,portmann2022security}. This means that even if the devices are potentially flawed or compromised, one can still certify the randomness of the outcomes based on observed correlations between the outcomes of measurements made on entangled quantum states.
The Entropy Accumulation Theorem (EAT) provides a framework for quantifying how much randomness can be certified from a sequence of measurements performed in a device-independent manner~\cite{arnon2018practical,dupuis2020entropy}. Specifically, it establishes conditions under which one can accumulate entropy over multiple rounds of measurements, leading to an overall increase in certified randomness. The EAT theorem outlines how subsequent measurements can contribute to an overall increase in entropy when certain conditions are met—primarily focusing on correlations that exceed classical limits often quantified using Bell inequalities~\cite{bell1964einstein,brunner2014bell}.

The paper is organized as follows. In Section~\ref{sec:Bell}, we present a selection of Bell inequalities used as certificates of randomness. We explore the criteria for choosing appropriate Bell expressions to ensure robust randomness certification, which is fundamental to the operation of Quantum Random Number Generators (QRNGs). Section~\ref{sec:EAT} delves into EAT and the certification parameters used in our analysis. We detail the theoretical framework that underpins the certification of randomness, highlighting key parameters such as the smoothing parameter ($\epsilon_{\text{s}}$) and its implications for randomness quality. In Section~\ref{sec:experiment}, we describe the experimental setup used to implement the QRNG protocol. We provide a detailed account of the components and configuration, including the sources of entangled photons and detection mechanisms. Section~\ref{sec:laserStrength} focuses on adjusting pump laser strength to optimize the generation rate of random numbers. We analyze the trade-offs between event rates and the quality of Bell violations, providing insights into the optimal operational conditions for the QRNG. In Section~\ref{sec:smoothTime}, we examine the impact of the smoothing parameter and generation time on the net generation rate. We discuss how different levels of the smoothing parameter affect the stability and reliability of the protocol, and the practical considerations for selecting an appropriate generation time. Section~\ref{sec:confAndTestRounds} investigates the role of confidence intervals and test rounds in the QRNG protocol. We explain the necessity of balancing the number of test rounds and the confidence level ($p_\Omega$) to ensure the robustness of the randomness certification process. In Section~\ref{sec:switchDelay}, we explore the impact of switch delay on the generation rate when changing measurement settings. We analyze how delays in the experimental setup can affect the efficiency of the QRNG and propose strategies to mitigate these effects. Finally, in Section~\ref{sec:discussion}, we provide a discussion and conclusions. We summarize the findings of our study, highlighting the key factors influencing the performance and security of QRNGs. We also suggest potential avenues for future research and improvements in QRNG technology.

\section{Bell certificates of randomness}
\label{sec:Bell}

Bell inequalities serve as a test for the predictions of quantum mechanics against those of classical physics, particularly local hidden variable theories. They were derived by John Bell in 1964 and are designed to assess whether the correlations observed between measurements on entangled particles can be explained by any deterministic theory that adheres to locality~\cite{bell1964einstein,horodecki2009quantum}.
The Tsirelson bound refers to the upper limit on the strength of correlations predicted by quantum mechanics when measuring entangled particles~\cite{cirel1980quantum}. While classical theories can reach at most a certain limit, the classical bound, quantum mechanics allows for stronger correlations than those permitted by classical physics but still respects this Tsirelson bound. The significance of this bound lies in its role in distinguishing between different types of correlations—those achievable through classical means versus those achievable through quantum entanglement.
The CHSH inequality is a specific form of Bell inequality formulated by Clauser, Horne, Shimony, and Holt in 1969~\cite{clauser1969proposed}. It involves two parties (Alice and Bob) each choosing between two measurement settings (A0, A1 for Alice and B0, B1 for Bob). If Bell inequality is violated it gives an opportunity to provide certification of security of random numbers even when devices are not trusted~\cite{pironio2010random}. There is on going research on finding the most suitable  Bell inequalities for this purpose~\cite{mironowicz2013robustness,wooltorton2022tight}.

We define the correlators $C(x,y)$ between Alice's results of the measurement $x$ and Bob's results of the measurement $y$ with the following equation:
\begin{equation}
	\begin{aligned}
		C(x,y) &\equiv P(0,0|x,y) + P(1,1|x,y) \\
		&- P(0,1|x,y) - P(1,0|x,y).
	\end{aligned}
\end{equation}
To certify the randomness, we utilized two Bell expressions, viz. CHSH, and one from a family introduced in~\cite{wooltorton2022tight} and defined with the following expression:
\begin{equation}
	\label{eq:Bell2}
	C(0,0) + 2.0126 \times C(0,1) + 2.0126 \times C(1,0) - 1.9754 \times C(1,1). % C(0,0) + 2.0126*C(0,1) + 2.0126*C(1,0) - 1.9754*C(1,1)
\end{equation}

\section{Entropy Accumulation and certification parameters}
\label{sec:EAT}

We utilize these two Bell expressions for quantum randomness certification with various parameters. In the work we follow the formulation of EAT provided in~\cite{brown2019framework}. A crucial tool needed to calculate the certified entropy with EAT is the min-trade-off function, providing a lower bound on the entropy in a single event for a given value of Bell violation.

The first parameter is the so-called smoothing parameter of randomness, $\epsilon_{\text{s}}$. This parameter specifies our requirement about the quality of the randomness, as further used by randomness extractors. The smaller the value of $\epsilon_{\text{s}}$ the better randomness.

The other parameter is the value of the violation of the Bell inequality. The higher the quality of the experiment, the higher the violation should be. On the other hand, in the experiment using the spontaneous parametric down-conversion (SPDC) process, a higher emission rate of entangled pairs of photons usually decreases the quality of the particular Bell violation. The number of detected pairs depends on the efficiency of detectors and transmission and coupling losses, and its rate is denoted as "events per second".

Any real-world device must provide results in a finite time to be valid. Finite data size effects are accounted for with EAT. In EAT, we need to set a generation time. The longer the generation time, the lower the negative impact of finite data size. However, for a device to be convenient, we need shorter generation times, as the device provides the generated randomness in bulks of data, i.e., a whole group of bits obtained during the generation time, and the user doesn't want to wait for them for too long.

For randomness generation, only a particular pair of settings $(x_0, y_0)$ of Alice and Bob are considered. To estimate the value of Bell violation in some rounds of the protocol, we need to randomly choose settings different from $(x_0, y_0)$. The rounds used for generation are called generation rounds, and those used for testing are called test rounds. The probability of test rounds is denoted by $\gamma$. This approach is called spot-checking. Every test round consumes public randomness for the random choice of settings $x$ and $y$; generation rounds don't consume additional randomness. The larger $\gamma$, the better the estimation of Bell violation, but more randomness is consumed, and fewer rounds are used for generating new randomness. In certification, we know the Bell violation only up to a given confidence interval $p_\Omega$, and we always need to assume the most pessimistic violation value to keep the protocol secure. Therefore, there is a non-trivial relation between the generation rate and the values of $\gamma$ and $p_\Omega$.

Since we need to change the optical components settings in the experimental setup for the test rounds, every change of settings usually introduces a certain delay, depending on the devices used. This delay additionally decreases the rate of events and, consequently, the rate of the protocol.

The parameters discussed above directly relate to the devices' physical workings. Apart from them, the formulae appearing in EAT contain an additional parameter, usually denoted as $\beta$, which is in the range $(0,1)$ and can be chosen arbitrarily.

\section{Experimental setup}
\label{sec:experiment}

\begin{figure}[htbp]
	\includegraphics[width=\columnwidth]{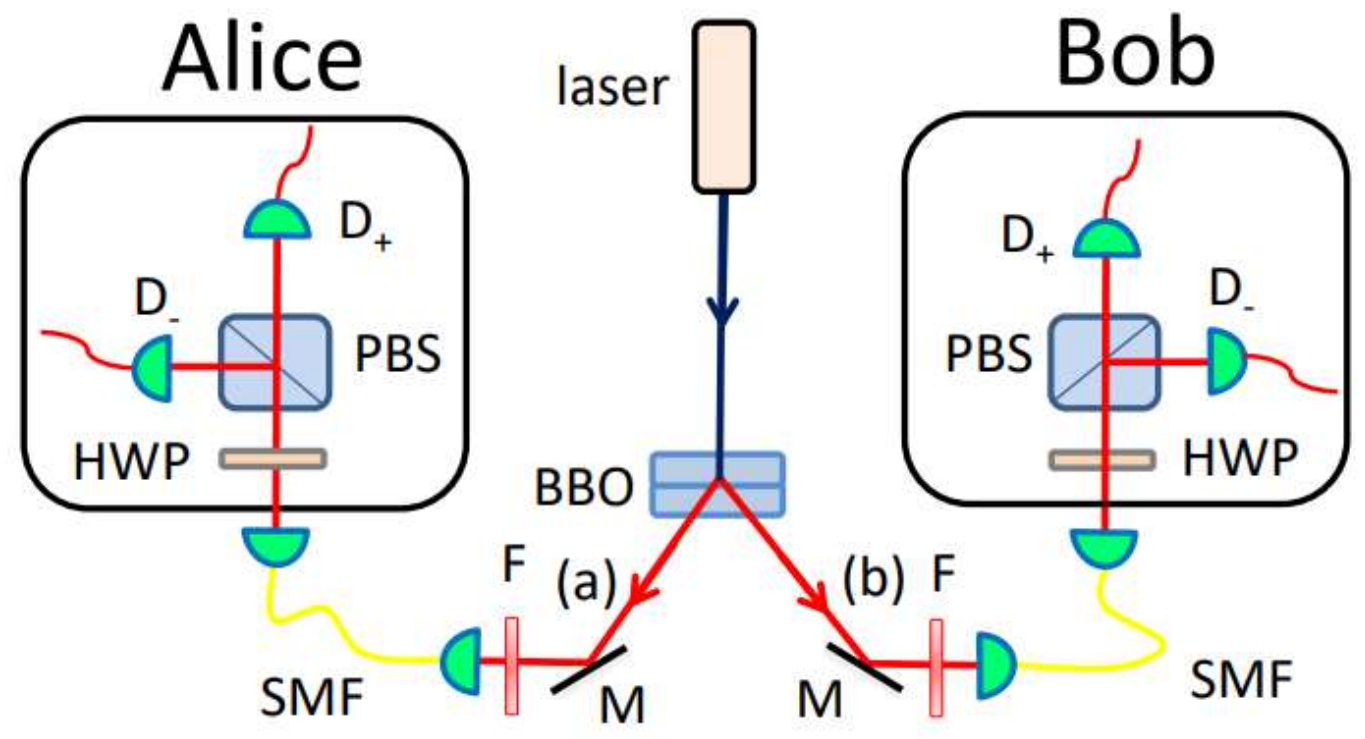}
	\caption{(Color online) Experimental setup. Entangled photon pairs are generated through the SPDC process. The emitted photons in modes  (a) and  (b) are coupled to single-mode fiber (SMF) and pass narrow filters (F). Each of the two stations’ measurements is composed of a half-wave plate (HWP), a polarization beam splitter (PBS), and single photon detectors ($D_{+}$ and  $D_{-}$). (See main text for details).}
	\label{fig:setup}
\end{figure}

The experimental setup is illustrated in Fig.~\ref{fig:setup}.
To generate the polarization entangled pair in state $\ket{\phi^{+}}=\frac{1}{\sqrt{2}}(\ket{HH}+\ket{VV})$, two 2 mm thick $\beta$ barium borate nonlinear crystals type 1 placed in interferometric configuration are pumped with a femtosecond in pulsed laser at wavelength of 390nm, to produce photon pairs emitted into two spatial modes (a) and (b) through the degenerate spontaneous parametric down-conversion process. The spatial, spectral, and temporal distinguishability between the down-converted photons is carefully removed by coupling to single-mode fiber, passed through narrow-bandwidth interference filters (F) and quartz wedges, respectively. The pump power of the femtosecond laser is tunable, allowing the photon pair generation rate to be set according to the experiment's requirements.
Alice and Bob's measurements are performed with the help of half-wave plates ($HWP_{A}, HWP_{B}$) oriented respectively $\theta_{A0}$ or $\theta_{A1}$, and $\theta_{B0}$ or $\theta_{B1}$. The polarization measurement was performed using PBS, and single-photon detectors (D) were placed in the two output modes of the PBS. Our detectors are actively quenched Si-avalanche photodiodes. All single-detection events were registered using a VHDL-programmed multichannel coincidence logic unit with a time coincidence window of 1.7 ns.
 
The photo pair emission rate was set for each measurement, and the corresponding measurement time was adjusted to keep the same number of 2-photon coincidence events for the measurement setting.
%This number of events is approximately 21 million for testing the inequality (1) and 10 million for testing the CHSH inequality.
At low rates, the multi-photon pair emissions are small, and therefore, the accidental events can be neglected. At higher rates, the accidental events will affect the correlations and the level of inequality violation. The error was estimated for each experiment by taking the standard deviation.

\section{Adjusting SPDC pump laser power for generation rate}
\label{sec:laserStrength}

The first decision when designing a QRNG is selecting the appropriate Bell inequality to use as the certificate. Next, we must determine how to adjust the SPDC pump laser power to achieve the optimal generation rate. This involves balancing the number of events per second with the Bell violation, as a higher laser power typically results in more events per second but a lower Bell violation, leading to less certified randomness per event. To estimate the optimal settings, we tested various event rates for the two Bell inequalities, CHSH and~\eqref{eq:Bell2}. We then calculated the asymptotic randomness generation rate for an infinite number of repetitions, thereby ignoring EAT's finite data size effects.

We computed the amount of certified entropy using two methods: the Navascues-Pironio-Acin (NPA) method~\cite{navascues2007bounding,navascues2008convergent}, which calculates a lower bound on certified min-entropy~\cite{chor1988unbiased,impagliazzo1989pseudo,konig2009operational,issa2017measuring}, and the Brown-Fawzi-Fawzi (BFF) method~\cite{brown2021device}, which calculates a lower bound on certified von Neumann entropy. Although the actual von Neumann entropy is always lower bound by min-entropy, the results from both methods may not align perfectly as they are based on different relaxation techniques. The numerical calculation we conducted using NCPOL2SDPA~\cite{wittek2015algorithm}, and MOSEK Solver~\cite{mosek21}.

Both NPA and BFF can be executed at various levels, forming a hierarchy. Higher levels offer better approximations but require more computational resources. We used level 2 of the hierarchy for NPA, which needed about a dozen seconds of computation. Higher levels, such as 3 and 4, did not improve the approximation significantly in our cases. For BFF, we also used level 2 of the hierarchy. Additionally, BFF involves another parameter: the number of nodes in the Gauss-Radau quadrature used in the approximation formulas. We considered using six and eight nodes. The computations took approximately 2.5 minutes for~\eqref{eq:Bell2} and 7 minutes for CHSH with six nodes. With eight nodes, the computation times increased to 3 minutes for~\eqref{eq:Bell2} and 7.5 minutes for CHSH, but the resulting estimates improved by only about 0.4\% and 0.8\%, respectively. The min-tradeoff function is obtained from the dual solutions of semi-definite optimization used in NPA and BFF with the method from sec.~3.4 of~\cite{mironowicz2024semi}. The min-tradeoff function has to be calculated once for each Bell violation value and doesn't need to be recalculated in further fine-tuning of protocol parameters.

\begin{table}[htbp]
	\begin{tabular}{|p{0.17\linewidth}|p{0.14\linewidth}|p{0.14\linewidth}|p{0.14\linewidth}|p{0.14\linewidth}|p{0.19\linewidth}|}
		\hline
		events per second & Bell value & $H_\text{min}$ & vNe Radau 6 & vNe Radau 8 & Asymptotic rate \\ \hline
		28000             & 4.95151    & 0              & 0           & 0           & 0               \\
		24000             & 5.00247    & 0.0098         & 0.0186      & 0.0186      & 446             \\
		20000             & 5.02311    & 0.1231         & 0.2234      & 0.2244      & 4488            \\
		16000             & 5.03036    & 0.1651         & 0.2953      & 0.2965      & 4744            \\
		12000             & 5.04098    & 0.2289         & 0.4007      & 0.4024      & 4829            \\
		8000              & 5.08671    & 0.5413         & 0.8545      & 0.858       & 6864            \\ \hline
	\end{tabular}
	\caption{Dependence of the certified min-entropy ($H_\text{min}$) and von~Neumann entropy (vNe) on the violation of the Bell inequality~\eqref{eq:Bell2} and the number of events per second. The values of $H_\text{min}$ are calculated with NPA level $2$. The values of von~Neumann entropy are calculated with approximations using Gauss-Radau quadrature with $6$ or $8$ nodes.}
	\label{tab:prop1_Bell}
\end{table}

\begin{table}[htbp]
	\begin{tabular}{|p{0.17\linewidth}|p{0.14\linewidth}|p{0.14\linewidth}|p{0.14\linewidth}|p{0.14\linewidth}|p{0.19\linewidth}|}
		\hline
		events per second & Bell value & $H_\text{min}$ & vNe Radau 6 & vNe Radau 8 & Asymptotic rate \\ \hline
		70000             & 2.65022    & 0.5198         & 0.8909      & 0.8964      & 62748           \\
		50000             & 2.67602    & 0.5638         & 0.9497      & 0.9574      & 47870           \\
		36000             & 2.70257    & 0.6148         & 1.0156      & 1.0239      & 36860           \\
		20000             & 2.71497    & 0.6411         & 1.0483      & 1.0566      & 21132           \\
		12000             & 2.73685    & 0.6925         & 1.1092      & 1.1177      & 13412           \\
		8000              & 2.74428    & 0.7117         & 1.1309      & 1.1397      & 9118            \\
		4000              & 2.76091    & 0.7591         & 1.1831      & 1.1917      & 4767            \\ \hline
	\end{tabular}
	\caption{Dependence of the certified min-entropy ($H_\text{min}$) and von~Neumann entropy (vNe) on the violation of the Bell inequality CHSH and the number of events per second. The values of $H_\text{min}$ are calculated with NPA level $2$. The values of von~Neumann entropy are calculated with approximations using Gauss-Radau quadrature with $6$ or $8$ nodes.}
	\label{tab:prop2_Bell}
\end{table}

The results are shown in Table~\ref{tab:prop1_Bell} for~\eqref{eq:Bell2} and Table~\ref{tab:prop2_Bell} for CHSH. Table~\ref{tab:prop1_Bell} shows that the best asymptotic rate for~\eqref{eq:Bell2} is achieved at 8000 events per second. Deviating from this event rate results in lower rates. Conversely, as shown in Table~\ref{tab:prop2_Bell}, the optimal asymptotic rate for CHSH occurs at the highest considered event rate, 70,000 events per second. We expect that further increasing the event rate will eventually cause a similar effect as seen in Table~\ref{tab:prop1_Bell}, where increasing the laser power beyond a certain point will start to decrease the asymptotic generation rate. For the remainder of this work, we will focus on the best asymptotic case, 70,000 events per second for CHSH.

%\section{Protection against detection efficiency loophole}
%
%Eberhard's approach provides a method to address the detection efficiency loophole in Bell experiments~\cite{eberhard1993background}. This loophole arises due to the possibility that some events, such as no-click events in detectors, may go undetected, leading to incomplete data and potentially biased results. To mitigate this issue, Eberhard proposed treating these no-click events as new types of measurement outcomes, thereby expanding the range of possible measurement results. By considering these undetected events as valid outcomes, Eberhard's method aims to provide a more comprehensive analysis of experimental data and reduce the impact of detection inefficiencies on the results of Bell tests.
%On the other hand, the assignment strategy, offers an alternative approach to address the detection efficiency loophole~\cite{wilms2008local,czechlewski2018influence}. In this strategy, instead of treating the no-click events as new types of outcomes, they are assigned one of the arbitrarily chosen values of the already existing types of outcomes. This assignment allows for a more straightforward analysis of the experimental data, as it preserves the structure of the measurement outcomes while accounting for the possibility of undetected events. While both approaches aim to mitigate the detection efficiency loophole, they differ in their treatment of undetected events and may have different implications for the interpretation of Bell test results.

\section{Smoothing parameter and generation time}
\label{sec:smoothTime}

As discussed earlier, one crucial parameter in QRNG is the smoothing parameter. The smaller this value, the higher the quality of the generated randomness, which is crucial for various applications. In our research, we explored different values of the smoothing parameter and observed that for the investigated protocol, reducing the smoothing parameter only slightly decreases the generation rate. Therefore, we opted for a much higher quality of randomness with $\epsilon_{\text{s}} = 1 \times 10^{-15}$.

Another significant parameter for practical applications of QRNG is generation time. Users typically prefer randomness to be generated as quickly as possible. However, while immediate generation is ideal, it is feasible to store larger segments (referred to as "single data chunks") of generated randomness and supply them on demand. This means that the generation time does not necessarily need to be in the milliseconds range; a time frame of several minutes is often acceptable. Nonetheless, minimizing the delay between generation and use is essential, as longer delays increase the risk of the randomness being compromised by an eavesdropper.

\begin{figure}[htbp]
	\centering
	\includegraphics[width=0.97\linewidth]{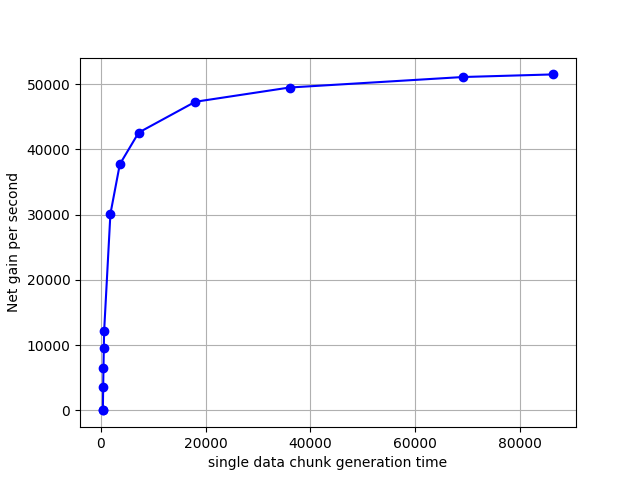}
	\caption{(Color online) The net generation rate for 70,000 events per second using the Bell inequality CHSH, with the smoothing parameter $\epsilon_{\text{s}} = 1 \times 10^{-15}$, depends significantly on the generation time for each data chunk. The minimum generation time required to achieve a positive rate is 6 minutes. The range of generation times explored extends up to 24 hours. It is important to note that the asymptotic generation rate reported in Table~\ref{tab:prop2_Bell} exceeds the values shown in the plot. This suggests that increasing the generation time can further mitigate the impact of finite data size effects, thereby enhancing the certified randomness rate. The data point at 69,120 seconds (19.2 hours) was specifically chosen to compare with the results presented in~\cite{liu2021device}, where this value was utilized.}
	\label{fig:generationtime}
\end{figure}

The relationship between the net generation rate and the generation time for a single data chunk is illustrated in Figure~\ref{fig:generationtime}. The data indicates that a minimum of 6 minutes is required to achieve a positive rate. The rate increases rapidly for generation times under 1 hour, reaching satisfactory levels at around 10 minutes. Consequently, we focus on these two generation times in our subsequent analysis. The data point at 19.2 hours demonstrates that our proposed protocol achieves over 50 kbps, significantly surpassing the 13,527 bits per second achieved with the same generation time in~\cite{liu2021device}.

\section{Confidence intervals and test rounds}
\label{sec:confAndTestRounds}

The next two critical elements in tuning a QRNG are the confidence intervals, expressing the probability that the protocol does not abort (denoted as $p_\Omega$), and the probability of test rounds (denoted as $\gamma$). Even if the average generation rate is high, a lower $p_\Omega$ means the generation process is less stable. In other words, if $p_\Omega$ is low, there is a relatively high probability that the Bell violation might not be significant enough to accept the generated chunk of randomness after the entire generation procedure. This can be a major drawback for the user experience, as the user might have to wait for the data to find out it cannot be used. To reduce the likelihood of the protocol aborting, one can either increase the tolerance for lower Bell violations or conduct more test rounds, thereby mitigating the risk that a lower Bell violation is due to statistical fluctuations rather than an eavesdropper's influence.

Thus, while test rounds can help stabilize the protocol's operation, they come with certain additional costs. The most explicit cost is the increased consumption of randomness required for these rounds. It is important to note that test rounds fail to generate new randomness and require additional randomness (2 bits for two settings) to choose the settings of Alice and Bob for estimating the Bell violation. Furthermore, selecting which rounds are test rounds also consumes randomness, although this is usually much smaller than the randomness needed during the test rounds themselves.

Alice and Bob's settings are chosen according to probability distributions $P_X$ and $P_Y$, respectively. Let $H_X$ and $H_Y$ denote the Shannon entropies of $P_X$ and $P_Y$. We have $H_X = H_Y = 1$ for uniform distribution of settings. The total randomness consumption for a single choice of settings for Alice and Bob in test rounds is denoted by $H_{XY} = H_X + H_Y$. The randomness consumed in selecting which rounds are test rounds is given by the binary entropy of $\gamma$, denoted as $H_2(\gamma)$. This randomness is consumed in every round, not just the test rounds. Therefore, the average consumption per round is calculated as:
\begin{equation}
    H_{XY} \times \gamma + H_2(\gamma).
\end{equation}

\begin{figure}[htbp]
	\centering
	\includegraphics[width=0.97\linewidth]{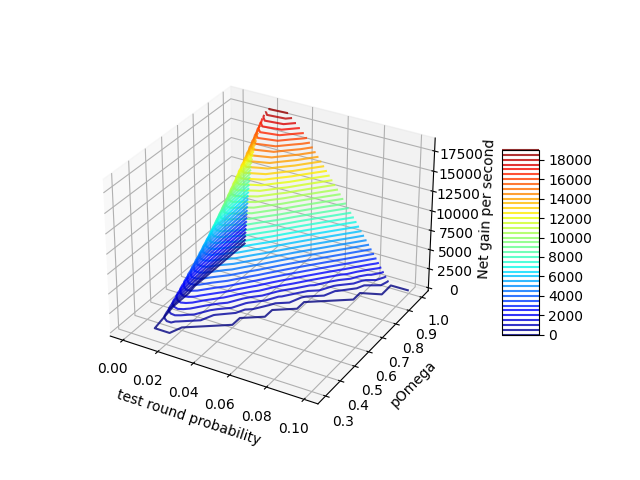}
	\caption{(Color online) Effect of the test round probability $\gamma$ and the protocol non-aborting probability $p_\Omega$ on the generation rate, for a single data chunk generation time of 10 minutes. This analysis assumes the smoothing parameter $\epsilon_{\text{s}} = 1 \times 10^{-15}$ and uses the Bell inequality CHSH at an event rate of 70,000 per second.}
	\label{fig:gammaandpomega10minutes}
\end{figure}

\begin{figure}[htbp]
	\centering
	\includegraphics[width=0.97\linewidth]{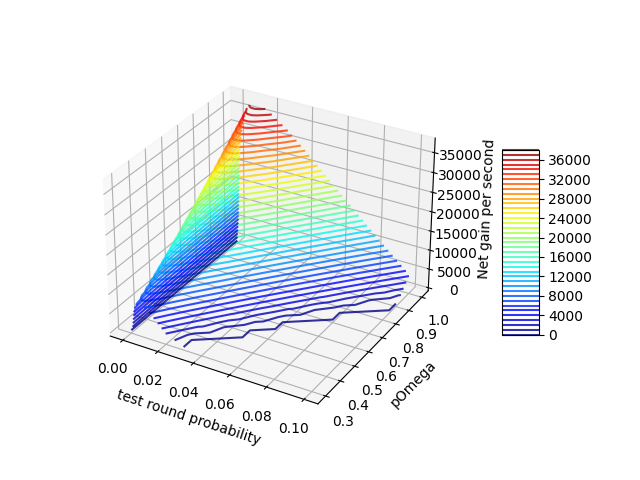}
	\caption{(Color online) Impact of the test round probability $\gamma$ and the protocol's non-aborting probability $p_\Omega$ on the generation rate, with a single data chunk generation time of 1 hour. The protocol parameters are consistent with those used in Figure~\ref{fig:gammaandpomega10minutes}.}
	\label{fig:gammaandpomega1hour}
\end{figure}

The generation rate's dependence on the test round probability $\gamma$ and the protocol's non-aborting probability $p_\Omega$ for single data chunk generation times of 10 minutes and 1 hour is shown in Figures~\ref{fig:gammaandpomega10minutes} and~\ref{fig:gammaandpomega1hour}, respectively. In both cases, it was optimal to use a $p_\Omega$ value close to 1. Thus, we focus on $p_\Omega = 0.9999$ for further analysis. To estimate the corresponding tolerance for the deviation of the observed Bell violation from the expected value, we used the estimation method from~\cite{pironio2010random}. The plots indicate that the optimal test round probability $\gamma$ decreases with longer generation times. It is important to note that this analysis assumes the test rounds' switching time is zero.

\begin{figure}[htbp]
	\centering
	\includegraphics[width=0.97\linewidth]{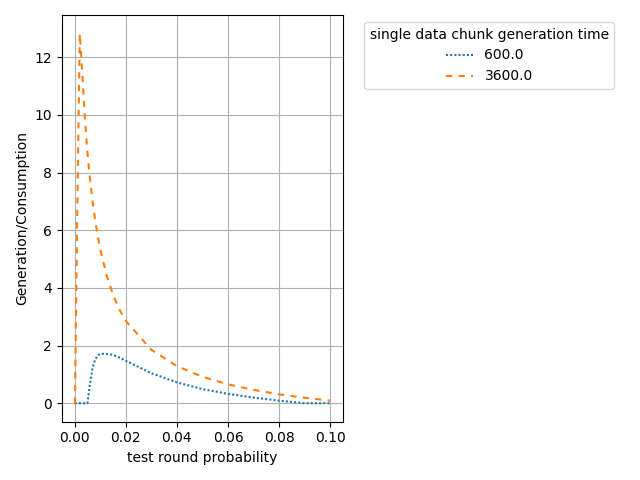}
	\caption{(Color online) The expansion rate, defined as the ratio of total net randomness generated to the total randomness consumed, for single data chunk generation times of 10 minutes and 1 hour. The analysis considers the smoothing parameter $\epsilon_{\text{s}} = 1 \times 10^{-15}$ and employs the Bell inequality CHSH with an event rate of 70,000 per second.}
	\label{fig:generationbyconsumption}
\end{figure}

The protocol must minimize the consumption of randomness, as an additional source of secure, unpredictable randomness must be provided for its operation. While the randomness does not need to be private, it should be unknown before the protocol runs. Otherwise, untrusted devices could potentially adapt to the known randomness, exploiting deterministic strategies to simulate a higher Bell violation.
A significant drawback of the approach described in~\cite{liu2021device} was its inefficiency in terms of randomness consumption. Specifically, generating $6.496 \times 10^9$ random bits required $6.233 \times 10^9$ bits, resulting in a net generation of only $2.63 \times 10^8$ bits. This led to an expansion rate of approximately $2.63 \times 10^8 / 6.233 \times 10^9 \approx 0.042$. In contrast, the present work reports a significantly better expansion rate, as demonstrated in Figure~\ref{fig:generationbyconsumption}.

\section{Impact of switch delay for measurement settings}
\label{sec:switchDelay}

\begin{figure}[htbp]
	\centering
	\includegraphics[width=0.97\linewidth]{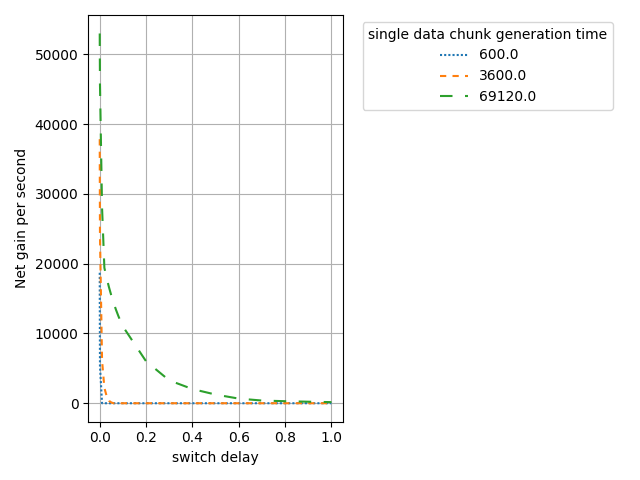}
	\caption{(Color online) Net generation rate as a function of switch delay for chunk generation times of 10 minutes, 1 hour, and 19.2 hours. For the 10-minute chunk duration, the net gain diminishes almost immediately with any switching delay. In the case of a 1-hour chunk, a small switch delay is tolerable but still results in a significant decrease in the generation rate. For a 19.2-hour chunk, the rate declines gradually with increasing delay. The analysis assumes the smoothing parameter $\epsilon_{\text{s}} = 1 \times 10^{-15}$ and uses the Bell inequality CHSH with an event rate of 70,000 per second.}
	\label{fig:switchdelay}
\end{figure}

The final parameter we investigate is the switching delay introduced by the time mechanical elements of the device are required to transition between various settings of Alice and Bob. Naturally, a longer switching delay has a more pronounced negative impact on the generation rate of the QRNG. To provide a comprehensive analysis, we consider single data chunk generation times of 10 minutes, 1 hour, and 19.2 hours, the latter included for reference as reported in~\cite{liu2021device}. The results are presented in Figure~\ref{fig:switchdelay}.

\begin{figure}[htbp]
	\centering
	\includegraphics[width=0.97\linewidth]{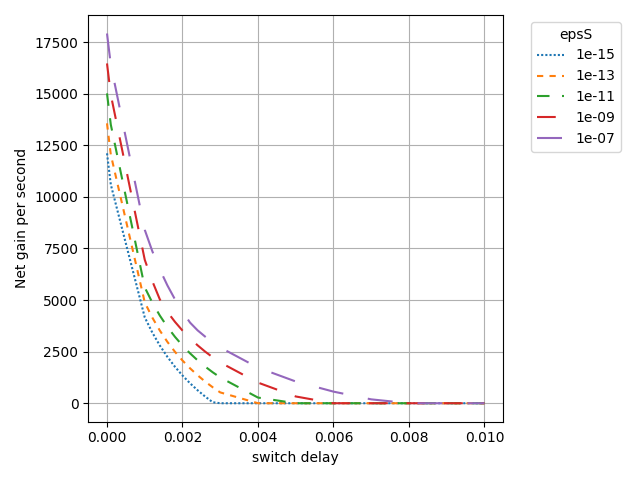}
	\caption{(Color online) Generation rate as a function of switch delay for a chunk generation time of 6 minutes and various values of the smoothing parameter $\epsilon_{\text{s}}$. The plot illustrates that increasing the smoothing parameter can slightly alleviate the negative impact of longer switch delays on the generation rate. We used the Bell inequality CHSH with an event rate of 70,000 per second.}
	\label{fig:switchdelayandepss600}
\end{figure}

To address the negative impact of switch delay on the generation rate, we observe that allowing a higher smoothing parameter can provide significant improvements. This effect is particularly noticeable for shorter chunk generation times. As demonstrated in Figure~\ref{fig:switchdelayandepss600}, for a chunk generation time of 10 minutes, the maximum switch delay that can be tolerated is 0.01s. By adjusting the smoothing parameter, we can mitigate some of the adverse effects of increased switch delays.

\section{Discussion and conclusions}
\label{sec:discussion}

In this work, we have explored the practical considerations and performance trade-offs in designing and implementing a Quantum Random Number Generator (QRNG) based on Bell inequality violations. Specifically, we analyzed the impact of various protocol parameters, such as the smoothing parameter ($\epsilon_{\text{s}}$), test round probability ($\gamma$), and switching delays, on the generation rate and randomness quality. Our findings provide insights for optimizing QRNGs in real-world applications.

The smoothing parameter is a crucial parameter in QRNGs, as it determines the quality and security of the generated randomness. A smaller $\epsilon_{\text{s}}$ implies a higher quality of randomness, which is critical for applications requiring high-security guarantees. However, maintaining a low smoothing parameter, such as $1 \times 10^{-15}$, is feasible and offers substantial improvements in randomness quality without severely compromising the generation rate.

Our investigation revealed that optimizing $\gamma$ and $p_\Omega$ is essential for balancing the trade-off between randomness consumption and generation rate. A higher $\gamma$ improves the accuracy of Bell violation estimation but reduces the net randomness generated due to increased randomness consumption in test rounds. Conversely, a high $p_\Omega$ ensures the protocol's stability by reducing the likelihood of protocol abortion, which is critical for practical usability. Our results indicate that maintaining $p_\Omega$ close to 1 is optimal, as it balances stability and efficiency.

Switching delays, which arise from the mechanical limitations of the experimental setup, significantly impact the generation rate, especially for shorter chunk generation times. Our study shows that even small delays can drastically reduce the net generation rate. However, increasing the smoothing parameter can partially compensate for the adverse effects of switching delays, providing a viable strategy for mitigating their impact.

Our results demonstrate an improvement over the QRNG design presented by Liu et al. (2021), which had a low expansion rate and limited net randomness generation. By optimizing the protocol parameters and considering longer chunk generation times, we achieved a higher expansion rate and a more robust performance under varying conditions, including switching delays. The choice of a high-quality smoothing parameter further distinguishes our approach, providing a better balance between security and efficiency.

Future research could focus on further optimizing the QRNG protocol by exploring additional parameters and configurations. In particular, investigating alternative Bell inequalities or quantum states may yield higher generation rates or better resilience to experimental imperfections. Additionally, developing faster and more precise switching mechanisms could reduce the impact of switching delays, enhancing the QRNG's overall performance.

Our work highlights the importance of carefully tuning QRNG protocol parameters to maximize performance and security. By considering the interplay between the smoothing parameter, test round probability, protocol stability, and switching delays, we provide a comprehensive framework for optimizing QRNGs. Our findings offer practical guidelines for designing robust and efficient QRNG systems capable of meeting the stringent requirements of secure applications.

\section*{Acknowledgements}

This work was supported by the Knut and Alice Wallenberg Foundation through the Wallenberg Centre for Quantum Technology (WACQT).

\bibliographystyle{ieeetr}

\end{document}